\documentclass{cernyrep}
\usepackage[T1]{fontenc}
\usepackage[bookmarks, colorlinks=true, linktoc=page, pdftex, linkcolor=black, citecolor=black, urlcolor=blue]{hyperref}

\usepackage{fancyhdr}
\usepackage{subfig} 
\usepackage[labelfont=bf]{caption}
\DeclareCaptionFont{tenpt}{\fontsize{10}{12}\selectfont}
\usepackage{url}
\usepackage{siunitx}

\usepackage{lineno} % Load the package
\fancyhfoffset{4 mm}
\fancypagestyle{ARTTITLE}{% 
\fancyhf{} % clear all header and footer fields
\lhead{\hfill CERN Accelerator School Proceedings --
{\it Intensity Limitations in Hadron Beams} --  Borovets, Bulgaria, 2025\hfill}
\lfoot{\hspace{3mm} Available online at \url{https://cas.web.cern.ch/previous-schools}}
\rfoot{\thepage\hspace*{3mm}}
 
}

\usepackage{varwidth}
\usepackage{xcolor}
\begin{document}

\title{Beam Loss Consequences}
\author{Giuseppe Lerner}
\institute{CERN, Geneva, Switzerland}

\begin{abstract}
The operation of high-energy and high-intensity particle accelerators inevitably leads to the loss of a fraction of beam particles, either through controlled processes or accidental events. This article builds on a first lecture on particle–matter interactions to review the main beam loss mechanisms in high-energy and high-intensity accelerators and their implications for safe and efficient operation. It discusses the resulting risks of equipment and material damage, radiation effects on electronics, and radiation-protection hazards. The~focus is on beam losses in hadron accelerators, with particular emphasis on the~Large Hadron Collider at CERN, while also addressing proposed future facilities such as the Future Circular Collider and muon colliders.
\end{abstract}
\maketitle

\thispagestyle{ARTTITLE}

\section{Introduction}

The unavoidable loss of a fraction of the beam during the operation of high-energy and high-intensity accelerators poses significant technical and safety challenges. As will be covered in this lecture, beam losses can arise from a range of dynamical processes that ultimately drive particles onto the machine aperture. These interactions typically generate radiation showers that deposit energy and may cause damage to accelerator elements and the surrounding environment. Having introduced the principles of particle-matter interactions in a first lecture~\cite{bib:LernerCAS2025interactions}, this article reviews the main beam loss mechanisms, focusing mainly on hadron accelerators, while also covering the challenges associated with beam losses at proposed facilities such as the Future Circular Collider (FCC)\cite{bib:FCCFeasibility} and muon colliders~\cite{bib:MuonCollider}. Specifically, accelerator beam losses are introduced in Section~\ref{sec:acc_beam_losses}, including a review of their sources at the Large Hadron Collider (LHC) at CERN~\cite{bib:LHC}. The article then examines the most critical consequences of beam losses in hadron accelerators, ranging from equipment and material damage (Section~\ref{sec:equipment_material_damage}) to radiation effects on electronics (Section~\ref{sec:r2e}) and radiation-protection aspects (Section~\ref{sec:rp}). Finally, beam loss considerations for the FCC and muon colliders are discussed in Section~\ref{sec:FCCee_mucol}.

\section{Accelerator beam losses}
\label{sec:acc_beam_losses}

\begin{figure}[t]
     \centering 
     \includegraphics[width=0.9\linewidth]{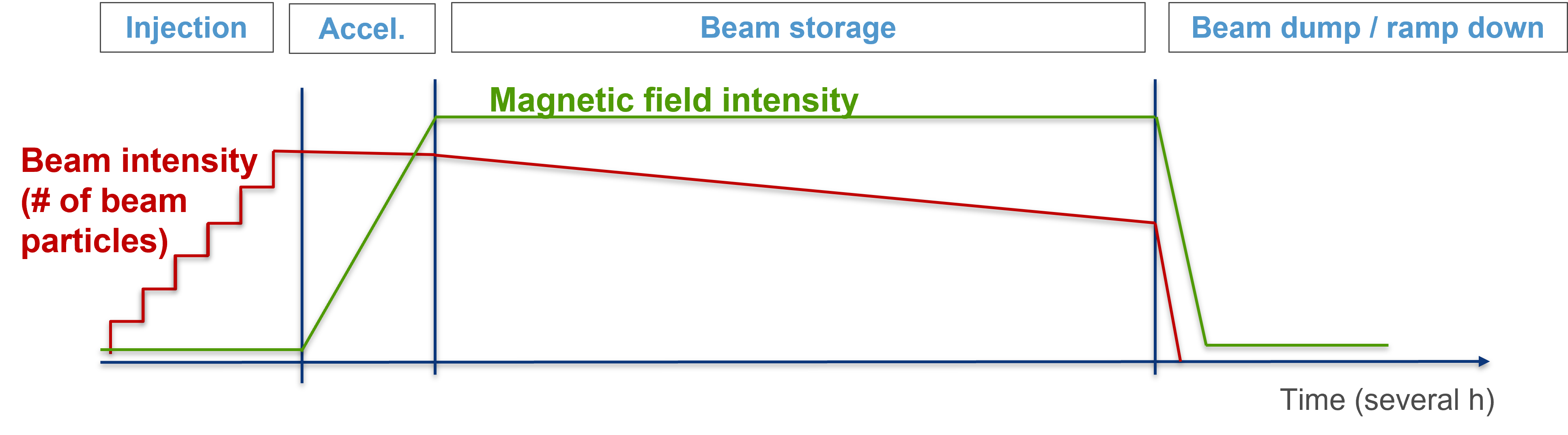}
     \caption{Schematic view of the operational cycle of a storage ring. \label{fig:storage_ring_operation}}
 \end{figure}

Beam losses in particle accelerators can originate from a wide range of processes, which may be either controlled (and in some cases deliberately induced) or accidental. In storage rings such as the Large Hadron Collider (LHC) at CERN, accelerator operation follows a well-defined cycle, schematically illustrated in Fig.~\ref{fig:storage_ring_operation}. The cycle begins with beam injection, during which the beams are introduced in successive steps (typically consisting of several bunches each) at an energy of $450$~GeV at the LHC. After injection is completed, the beams are accelerated by ramping up the magnetic fields of the bending dipoles, reaching the top energy of $6.8$~TeV during the ongoing LHC Run~$3$ ($2021\text{-}26$), slightly below the~design energy of $7$~TeV. This is followed by the main storage phase, during which the beams are brought into collision and maintained for several hours to deliver physics events to the experiments. The~operational cycle concludes with a beam dump, which may be either scheduled or triggered by an~abnormal condition, and the subsequent ramp down of the magnetic fields in preparation for the next cycle. At the LHC, in the absence of issues causing additional downtime, the turnaround time between two cycles (from a beam dump to the start of collisions in the subsequent cycle) is typically of the order of $2$~hours. Consequently, frequent beam dumps significantly reduce the effective time available for physics data taking and therefore have a substantial impact on overall machine performance.

A primary source of beam losses at the LHC arises from \textbf{inelastic proton-proton collisions} occurring at the four Interaction Points (IPs), which host the ATLAS~\cite{bib:ATLAS}, CMS~\cite{bib:CMS}, ALICE~\cite{bib:ALICE}, and LHCb~\cite{bib:LHCb} experiments. The number of inelastic collisions over a given operational period can be computed as:
\begin{equation}
N_{\mathrm{coll\text{-}inel}} = \sigma_{\mathrm{inel}} \times \mathcal{L}_{\mathrm{int}},
\end{equation}
where $\sigma_{\mathrm{inel}}$ is the inelastic proton–proton cross section, approximately $80$~mb at centre-of-mass energies of $13$--$14$~TeV. The quantity $\mathcal{L}_{\mathrm{int}}$, known as the integrated luminosity, is the reference figure of merit that quantifies the amount of data delivered to the experiments, providing the proportionality between the cross section of any process and the corresponding number of events. At the LHC, the annual $\mathcal{L}_{\mathrm{int}}$ reached approximately $120$~fb$^{-1}$ (in units of inverse femtobarn, where $1$~b~$=10^{24}$~cm$^2$) during Run~$3$, and it is expected to rise to $250\text{-}300$~fb$^{-1}$ per year following the High-Luminosity LHC (HL-LHC) upgrade~\cite{bib:HLLHC-TDR}. Similarly, one can define an instantaneous luminosity $\mathcal{L}_{\mathrm{int}}$, linked to the collision rate.

\begin{figure}[ht]
    \centering 
    \subfloat[\label{fig:CMS_collisions}]{
    \includegraphics[width=0.55\linewidth]{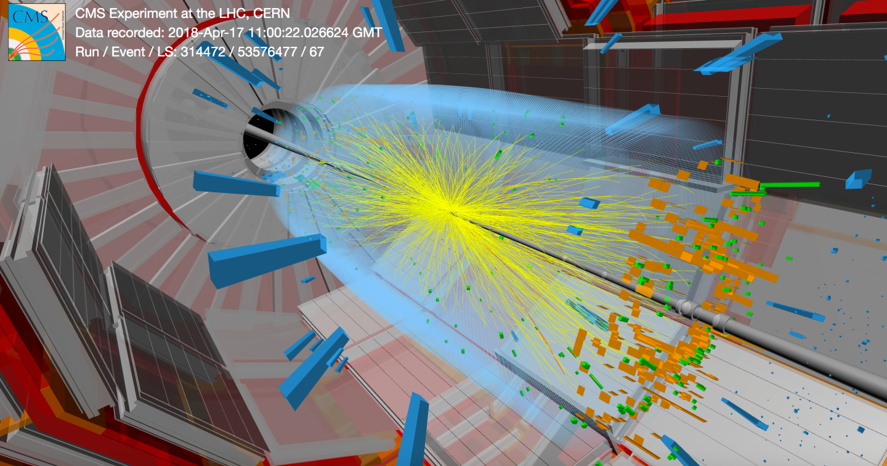}}
    \subfloat[\label{fig:LHC_particles_per_collision}]{
    \includegraphics[width=0.42\linewidth]{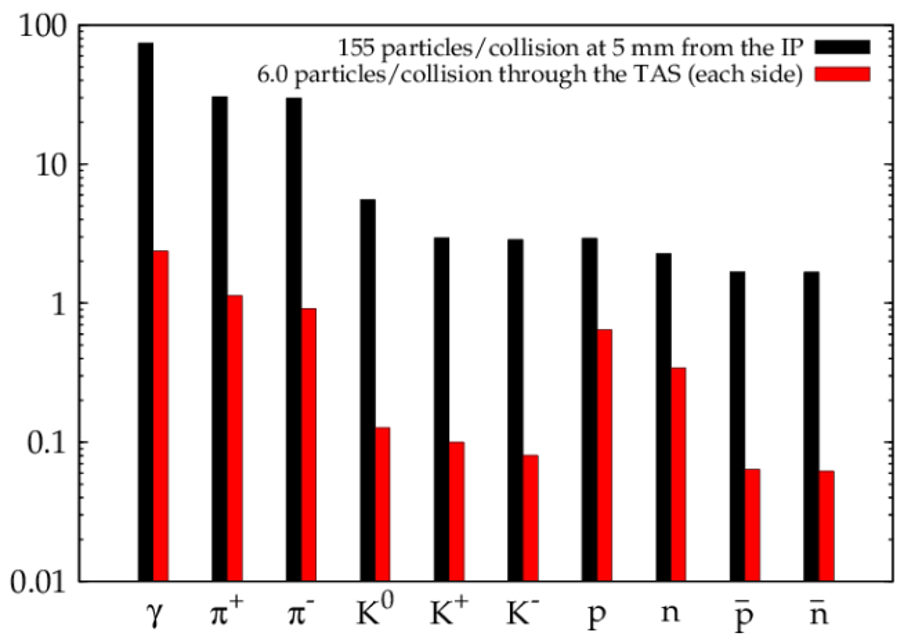}} 
    \caption{(a) Visualization of inelastic proton-proton collisions in the CMS experiment and (b) FLUKA-simulated breakdown of the average particle yield produced by single such collisions at $5$~mm from the interaction point (black) and at the exit of each $60$~mm TAS aperture (red), taken from Ref.~\cite{bib:HLLHC-TDR}. }
    \label{fig:LHC_collisions}
\end{figure}

A visual display of inelastic proton-proton collisions in the CMS experiment is shown in Fig.~\ref{fig:LHC_collisions}, along with the average number of particles per collision simulated with the FLUKA code~\cite{bib:FLUKA_website,bib:Ahdida2022,bib:Battistoni2015} at $5$~mm from the IP and reaching the LHC tunnel in the forward direction (beyond an absorber named TAS). The simulation results, described in detail in~\cite{bib:HLLHC-TDR}, indicate that only $\approx5\%$ of the particles produced in the collisions reach the LHC machine, but they also find that these few particles carry $\approx 80\%$ of the~power. At the HL-LHC, referring to a nominal instantaneous luminosity of $5\cdot10^{34}$~cm$^{-2}$~s~$^{-1}$, roughly $3.8$~kW of power are transported into the LHC tunnel on each side of the main IPs, causing radiation showers that propagate over hundreds of metres in the tunnel.

\begin{figure}[ht]
    \centering 
%    \subfloat[\label{fig:LHC_collimation_principle}]{
%    \includegraphics[width=0.25\linewidth]{Figures/LHC_collimation_principle.png}}
    \subfloat[\label{fig:LHC_collimation_system}]{
    \includegraphics[width=0.5\linewidth]{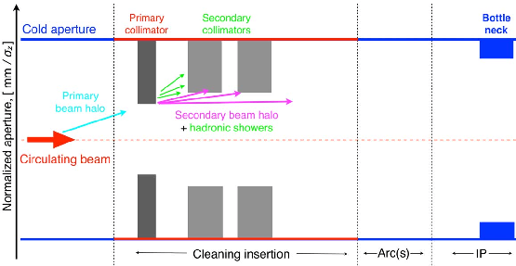}}
    \subfloat[\label{fig:LHC_beamgas_scheme}]{
    \includegraphics[width=0.41\linewidth]{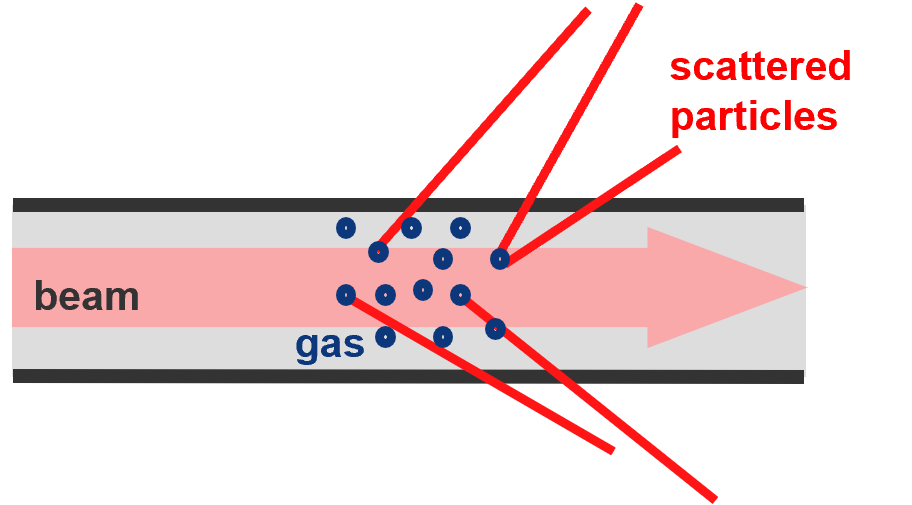}} 
    \caption{Schematic views of (a) the LHC multi-stage collimation system (from Ref.~\cite{bib:Redaelli2016}) and (b) beam-gas interactions in the LHC vacuum chamber. }
    \label{fig:LHC_collimation_beamgas}
\end{figure}

In high-energy, high-intensity accelerators such as the LHC, beam losses also occur in the \textbf{collimation system}, which is specifically designed to intercept unavoidable losses in a controlled manner~\cite{bib:Redaelli2016}. The LHC collimation system primarily targets the transverse beam halo using a multi-stage hierarchy, as illustrated in Fig.~\ref{fig:LHC_collimation_system}, although collimators are also deployed for other purposes (e.g., machine protection against accidental loss scenarios, and more). Primary collimators with the smallest normalized transverse apertures intercept halo protons, generating particle showers that are subsequently absorbed by secondary and tertiary collimators with progressively larger apertures. The collimator settings are deliberately tighter than the aperture at sensitive machine locations (e.g., near the IPs), ensuring that halo particles produced by multi-turn mechanisms are safely intercepted at predefined locations. Generally speaking, any interception of primary or secondary particles by a collimator (or by an aperture bottleneck) inevitably results in the development of radiation showers in the tunnel.

\textbf{Beam-gas interactions} in the vacuum chamber represent an additional important source of beam losses at accelerators such as the LHC. These losses are driven by inelastic nuclear collisions between the circulating protons and gas molecules, and their rate is proportional to the local gas density profile, the inelastic proton–gas cross section, and the beam intensity and revolution frequency (i.e., the beam current). Such interactions may occur either with the residual gas in the vacuum chamber or with gas deliberately injected by dedicated beam instrumentation exploiting beam–gas collisions, such as the beam gas curtain (BGC)~\cite{bib:BGC2024}. Beam–residual gas interactions typically dominate the radiation levels in the~LHC arcs, where other loss mechanisms are largely absent~\cite{bib:Bilko2023}. By contrast, in regions where collision-driven losses or collimation losses are present, beam–gas interactions are generally minor contributors to the~overall radiation field.

A simplified example of an LHC‑like radiation shower, generated by the loss of an injection‑energy proton in aluminum, is presented in Ref.~\cite{bib:LernerCAS2025interactions}. The impact of beam losses on accelerator components, electronics, and on radiation safety will be discussed in the following paragraph, where a central question is how to define observables associated with radiation fields to quantify the resulting damage.

\section{Equipment damage at hadron accelerators}
\label{sec:equipment_material_damage}

\begin{figure}[ht]
    \centering 
    \subfloat[\label{fig:energy_deposition_heating}]{
    \includegraphics[width=0.37\linewidth]{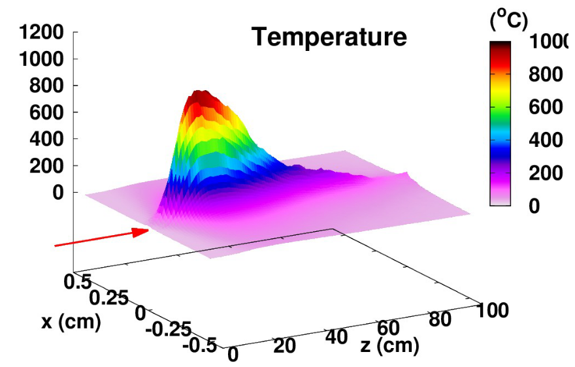}}
    \subfloat[\label{fig:SPS_accidental_loss}]{
    \includegraphics[width=0.615\linewidth]{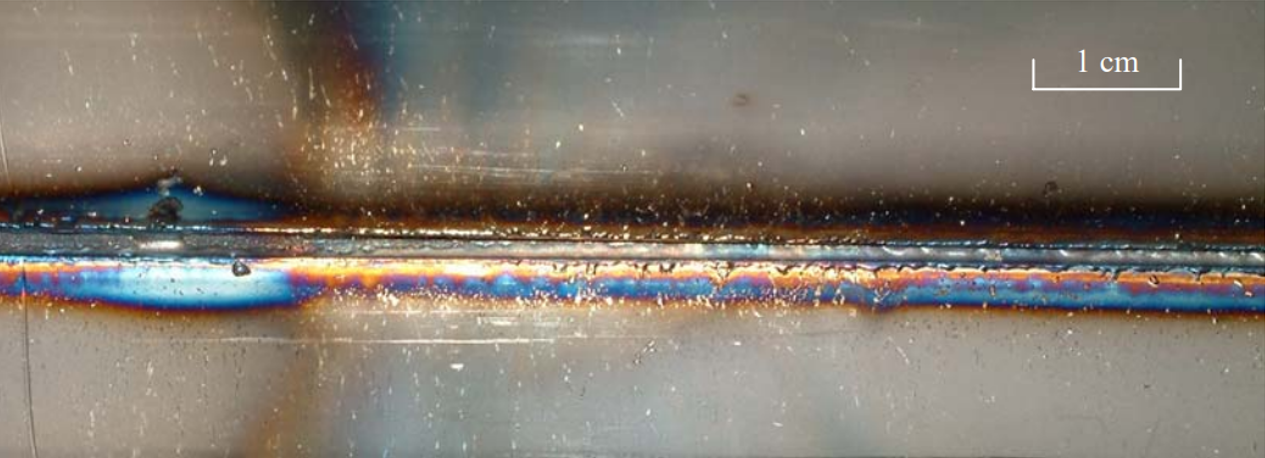}} 
    \caption{(a) Temperature increase map in a copper block caused by the energy deposited by a single $7$-TeV LHC bunch ($1.15\cdot10^{11}$ protons with a $0.3\times0.3$~mm spot size) simulated with FLUKA. (b) Damage to the vacuum chamber of the TT40 transfer line from the SPS to the LHC, caused by the accidental loss of a full LHC injection batch of $288$~bunches with $3.4\cdot10^{13}$ $450$-GeV protons in 2004 (figure from Ref.~\cite{bib:Goddard825806}). }
    \label{fig:equipment_material_damage_1}
\end{figure}

As already introduced in~\cite{bib:LernerCAS2025interactions} and in the above paragraphs, beam losses can lead to severe equipment damage through the development of intense radiation showers in matter. A key observable characterising these showers is the energy deposition (or power deposition when expressed per unit time), which represents the amount of energy transferred to the impacted materials, leading to a local temperature increase determined by the material's heat capacity. As an example, Fig.~\ref{fig:energy_deposition_heating} illustrates the effect of the energy deposited in copper by a single $7$-TeV LHC bunch consisting of $1.15\cdot10^{11}$~protons (a typical bunch intensity used in Run~$2$, also corresponding to the original LHC design), which produces a temperature increase that can reach \SI{1000}{\degreeCelsius}, close to the copper melting point of \SI{1085}{\degreeCelsius}. Real-world consequences of a similar impact were observed in 2004 at the CERN SPS, when an incident occurred during the extraction of an LHC-like beam ($3.4\cdot10^{13}$ protons at $450$~GeV, with a stored energy around $2.5$~MJ) causing major damage to the vacuum chamber, as shown in Fig.~\ref{fig:SPS_accidental_loss}. As a consequence, a full magnet replacement was required, leading to weeks of downtime~\cite{bib:Goddard825806}.

\begin{figure}[ht]
    \centering 
    \subfloat[\label{fig:LHC_BLMs}]{
    \includegraphics[width=0.6845\linewidth]{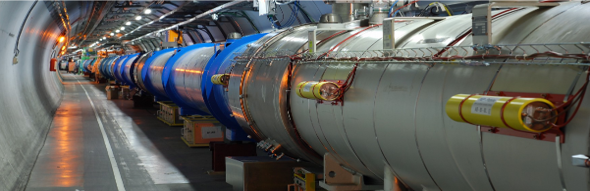}}
    \subfloat[\label{fig:LHC_RadMons}]{
    \includegraphics[width=0.3025\linewidth]{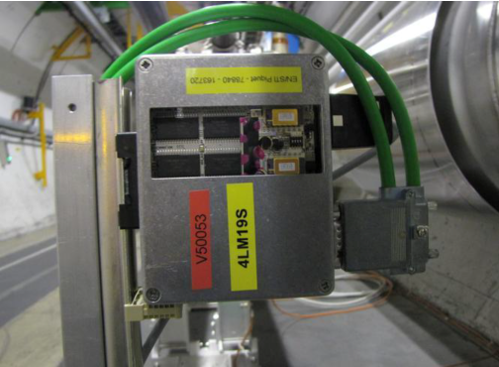}} 
    \caption{(a) LHC tunnel with Beam Loss Monitors (BLMs) on the cryostats and (b) a RadMon detector. }
    \label{fig:LHC_radiation_monitors}
\end{figure}

Even when macroscopic damage is avoided, energy deposition can have critical consequences in superconducting magnets. At LHC dipoles the temperature increase margin to maintain superconductivity can be as low as $\approx\!2$K, which can be triggered by instantaneous energy depositions of the order of mJ/cm$^3$~\cite{bib:Schoerling2023QuenchSlides}. Such quenches can require long recovery time (or even cause permanent damage), motivating the implementation of dedicated machine protection systems, including Beam Loss Monitors (BLMs, see Fig.~\ref{fig:LHC_BLMs}), which are ionisation chambers distributed around the LHC to characterise loss patterns and trigger a beam dump if pre-set thresholds are exceeded~\cite{bib:Holzer2005BLM}. Beyond immediate heating effects, energy deposition also manifests as Total Ionising Dose (TID), defined as the deposited energy per unit mass:
\begin{equation}
\label{eqn:TID_def}
    \mathrm{TID} = \mathrm{E}_\mathrm{dep}/\mathrm{M}, \quad \text{expressed in Gy} \stackrel{\text{def}}{=} \mathrm{J}/\mathrm{kg,}
\end{equation}
which can lead to the progressive degradation of exposed equipment and materials~\cite{bib:RadiationToMaterials2023}. In particular, TID-induced damage at the LHC can affect magnet coils (especially their insulation), as well as cables, lubricants, and electronics, as will be further described in the next paragraph. At the LHC, following the decision to extend Run~$3$, dedicated measures (including optics and crossing plane variations) have been implemented to ensure that the peak dose in the superconducting coils of quadrupole magnets near ATLAS and CMS does not exceed $30$~MGy~\cite{bib:LHCTripletTaskForce}.

Finally, Non-Ionising Energy Loss (NIEL) contributes to displacement damage, quantified by the~Displacements Per Atom (DPA), which measures the fraction of lattice atoms displaced from their equilibrium positions. This effect is especially relevant for silicon devices and superconducting materials, and it links accelerator radiation studies to broader research fields, such as fusion magnet technology.

\section{Radiation to Electronics (R2E) at hadron accelerators}
\label{sec:r2e}

Electronic systems at the LHC consist of up to thousands of distributed units, typically Commercial Off The Shelf (COTS) components, which may be exposed to varying levels of radiation during operation. Examples include power converters and their controls~\cite{bib:Uznanski2013REDW}, the Quench Protection System (QPS) that safeguards superconducting equipment from quenches~\cite{bib:Bitterling2016JINST}, and numerous other systems such as vacuum, cryogenics, and beam instrumentation. The racks housing these systems are either installed in the~tunnel, minimizing cabling distances, or in nearby shielded alcoves to mitigate radiation exposure. Radiation effects on electronics can be classified into cumulative and stochastic effects. \textbf{Cumulative effects}, such as lifetime degradation from Total Ionising Dose (TID) (Eq.~(\ref{eqn:TID_def})) and Displacement Damage (DD), have been introduced in Section~\ref{sec:equipment_material_damage} for materials, and they apply similarly to electronics. For DD, a~commonly used quantity for electronic devices is the Silicon 1-MeV equivalent fluence~\cite{bib:ASTM_E722}. 

\begin{figure}[ht]
    \centering 
    \includegraphics[width=0.65\linewidth]{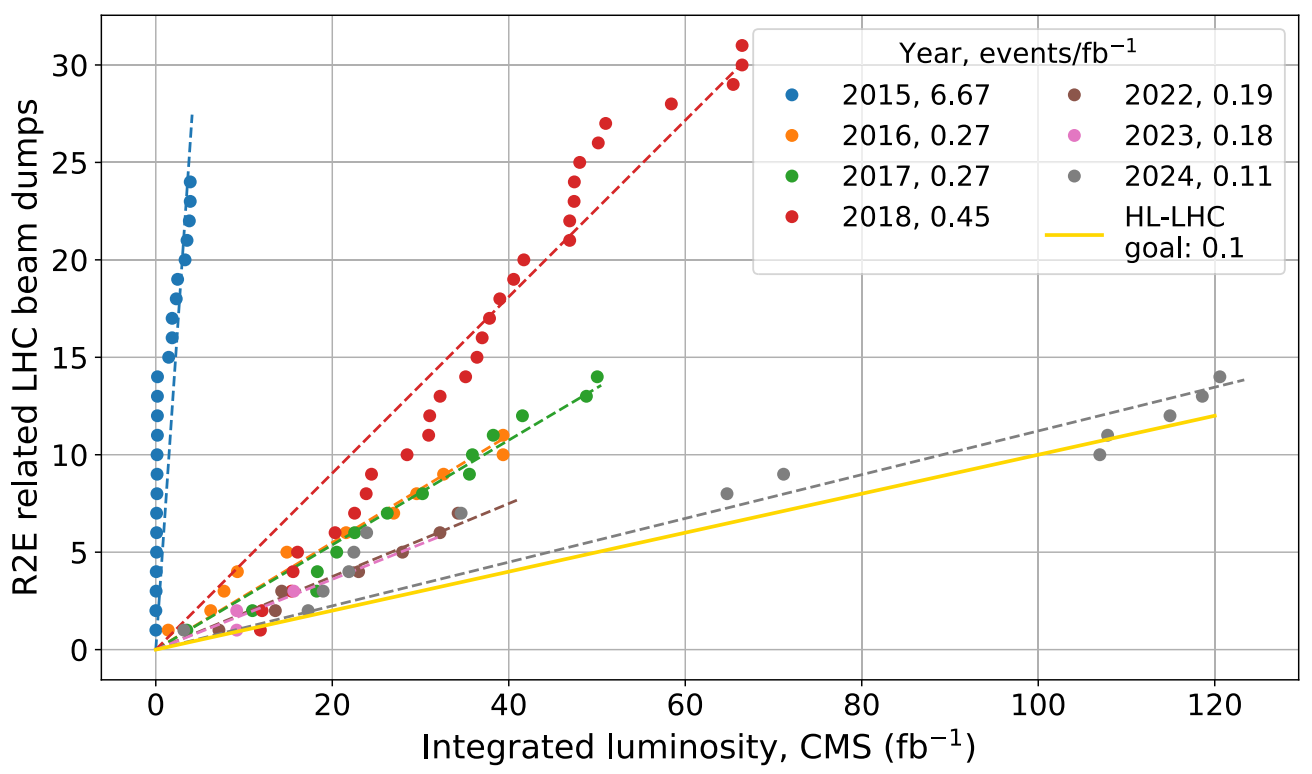}
    \caption{Number of radiation-induced LHC beam dumps during proton operation, as a function of the integrated luminosity delivered to the CMS experiment (from Ref.~\cite{bib:Soderstrom2025TNS}). }
    \label{fig:LHC_dumps_per_fb-1}
\end{figure}

Stochastic effects, commonly referred to as \textbf{Single Event Effects (SEEs)}, can induce failures in electronic systems at any time under exposure to a given radiation flux, independently of the device irradiation history. SEEs must be carefully evaluated for their impact on LHC availability, as even non-destructive failures may significantly disrupt the operation of critical systems during a run, forcing premature beam dumps during LHC cycles (Fig.~\ref{fig:storage_ring_operation}). A commonly used figure of merit for the LHC is the~number of R2E-induced beam dumps per unit integrated luminosity (fb$^{-1}$), i.e., the slope of the~curves shown in Fig.~\ref{fig:LHC_dumps_per_fb-1}, where the number of LHC beam dumps is plotted as a function of the annual integrated luminosity delivered to the CMS experiment~\cite{bib:Soderstrom2025TNS}; this metric has improved substantially since early LHC operation, approaching the ambitious target of 0.1 dumps per fb$^{-1}$ for the~HL-LHC upgrade.

\begin{figure}[t]
    \centering 
    \includegraphics[width=0.8\linewidth]{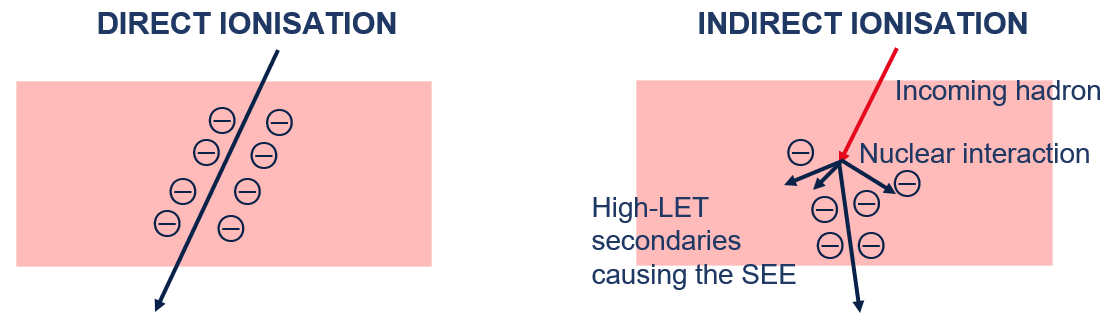}
    \caption{Schematic illustration of SEE generation mechanisms: (left) direct ionisation by a charged particle; (right) indirect ionisation following a nuclear interaction induced by an incident hadron. }
    \label{fig:direct_indirect_ionisation_SEEs}
\end{figure}

Physically, SEEs in semiconductor devices occur when incident particles generate charge carriers in the device’s small active volumes, as illustrated in Fig.~\ref{fig:direct_indirect_ionisation_SEEs}. While direct ionisation by high-LET particles (specifically heavy ions) is negligible at the LHC because such ions are absent from hadronic showers, SEEs there primarily arise from indirect ionisation, in which hadrons or photons induce nuclear reactions whose secondary particles deposit enough charge to trigger the effect. In general, the SEE rate can be calculated as
\begin{equation}
\label{eqn:def_SEEs_general}
N_{\rm SEE} = \Phi \cdot \sigma_{\rm SEE},
\end{equation}
where $\Phi$ is the particle fluence and $\sigma_{\rm SEE}$ is the device-specific SEE cross section. The fluence represents the density of particle track lengths within a volume (in units of cm$^{-2}$), while $\sigma_{\rm SEE}$ (in cm$^2$) characterises the susceptibility of a given electronic component to SEEs, using the same formalism of particle-matter interaction physics.

\begin{figure}[ht]
    \centering 
    \subfloat[\label{fig:LHC_spectrum_for_SEEs}]{
    \includegraphics[width=0.51\linewidth]{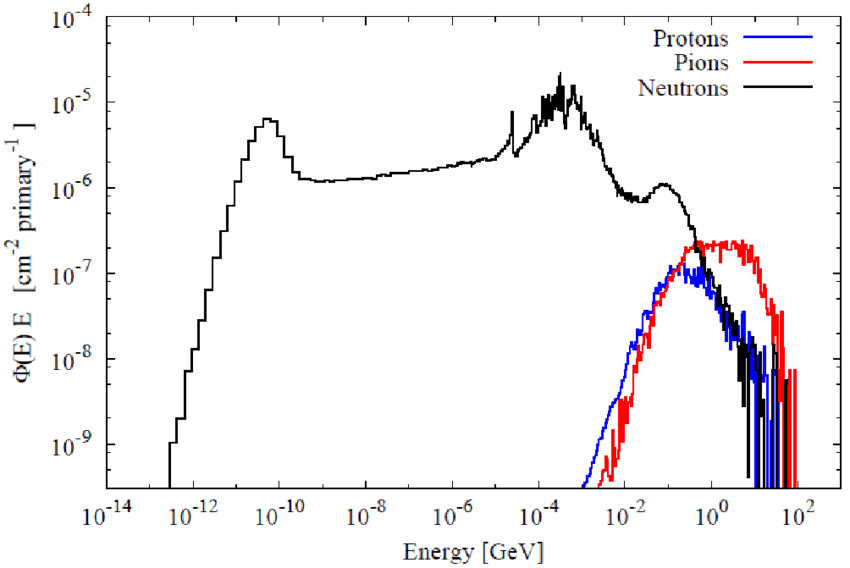}}
    \subfloat[\label{fig:SEE_xsec}]{
    \includegraphics[width=0.422\linewidth]{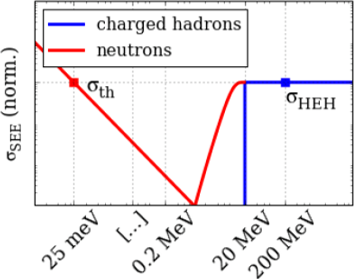}}
    \caption{(a) Energy spectrum of protons, pions, and neutrons in the LHC tunnel (from Ref.~\cite{bib:SpiezaRADMON}) and (b) cross section of SEEs induced via indirect ionisation as a function of the primary particle energy. }
    \label{fig:SEE_cross_section_and_LHC_spectrum}
\end{figure}

The primary particles inducing SEEs via indirect ionisation are neutrons and charged hadrons. Their energy spectra in the LHC tunnel, along with the SEE cross section as a function of energy, are shown in Fig.~\ref{fig:SEE_cross_section_and_LHC_spectrum}. Here, the cross section trend reflects the high-energy hadron (HEH) approximation: it is assumed constant for hadrons above 20~MeV, and null below, with neutrons in the intermediate 0.2–20~MeV range contributing according to a Weibull parametrisation~\cite{bib:cecchettoNeutronSER}:
\begin{equation}
\label{eqn:weibull}
    \sigma_{\rm SEE}(E) = \sigma_{\rm sat}\times w(E) = \sigma_{\rm sat} \times
\left[1 - \exp\!\left(-\left(\dfrac{E - E_{\rm th}}{W}\right)^{s}\right)\right],
\end{equation} 
where the parameters $\sigma_{\rm sat}$, $E_{\rm th}$, $W$, and $s$ are device-specific. A separate contribution arises from thermal neutrons through neutron capture reactions. Accordingly, the SEE rate calculation in Eq.~(\ref{eqn:def_SEEs_general}) can be split into two terms:
\begin{equation}
\label{eqn:def_SEEs_HEH_and_thermals}
    N_{\rm SEE} = \Phi_{\rm HEH-eq} \, \sigma_{\rm SEE}^{\rm HEH-eq} + \Phi_{\rm thn} \, \sigma_{\rm SEE}^{\rm thn},
\end{equation}
where the High Energy Hadron equivalent (HEH-eq) term embeds the contribution from hadrons with E$>20$~MeV and neutrons in the intermediate energy range. 

When qualifying electronics, the SEE cross sections $\sigma_{\rm SEE}^{\rm HEH}$ and $\sigma_{\rm SEE}^{\rm thn}$ appearing in Eq.~(\ref{eqn:def_SEEs_HEH_and_thermals}) are typically measured under simplified irradiation conditions, obtaining their values at specific energies (as shown in Fig.~\ref{fig:SEE_cross_section_and_LHC_spectrum}). The energy dependence of the SEE response can be absorbed in the definition of energy-integrated equivalent fluences with appropriate energy-dependent weighting functions, such that the High-Energy Hadron equivalent fluence, $\Phi_{\rm HEH\text{-}eq}$, is defined as:
\begin{equation}
\label{eqn:HEH-eq_def}
    \Phi_{\rm{HEH\text{-}eq}} = \int _{0.2~\rm {MeV}}^{20~\rm{MeV}} w(E) \cdot \frac {d \Phi _{n}\left ({E}\right)}{dE} \, dE~  \,\, + \int _{20~\rm{MeV}}^{+\infty } \frac {d \Phi _{\text {HEH}}\left ({E}\right)}{dE}\, dE
\end{equation}
with separate terms for neutrons in the intermediate energy range weighted by the Weibull (Eq.~(\ref{eqn:weibull})) and hadrons above 20~MeV. The thermal neutron equivalent fluence, $\Phi_{\rm thn}$, is defined analogously using weighting factors proportional to the inverse neutron velocity, i.e., scaling as $E^{-1/2}$. 

\begin{figure}[ht]
    \centering 
    \subfloat[\label{fig:FLUKA_BLMs_P1}]{
    \includegraphics[width=0.48\linewidth]{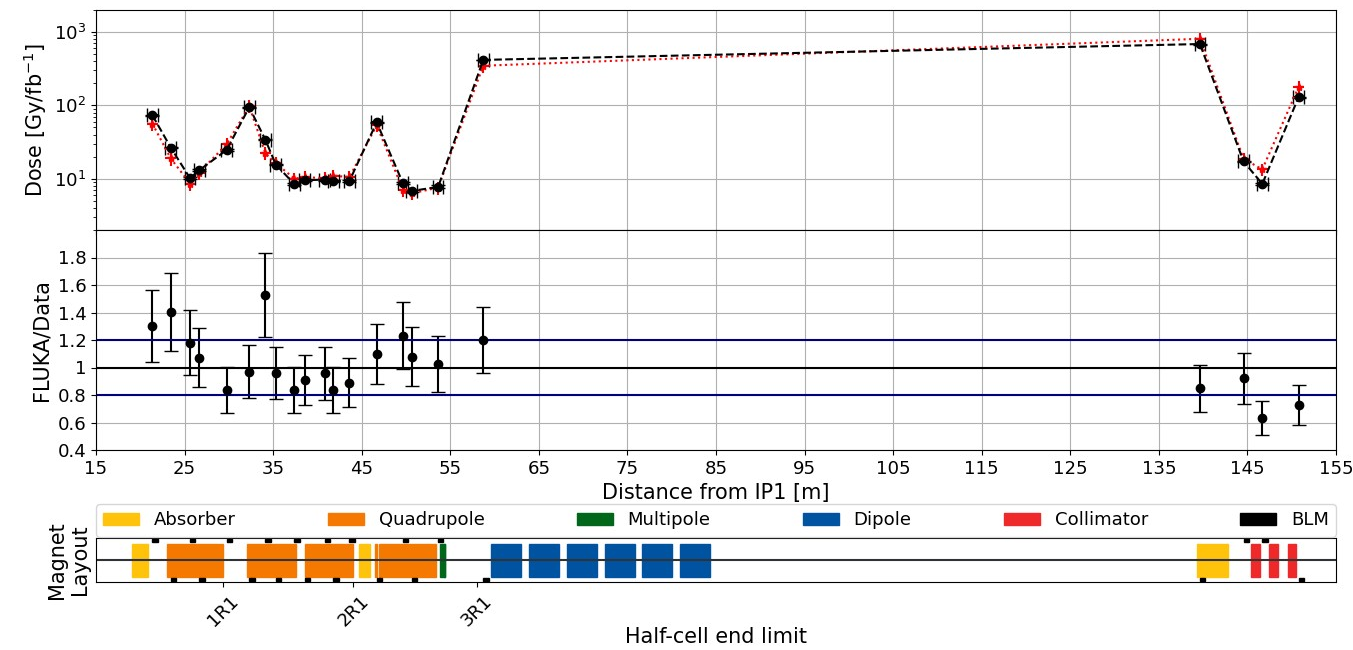}}
    \subfloat[\label{fig:FLUKA_map_RR17}]{
    \includegraphics[width=0.50\linewidth]{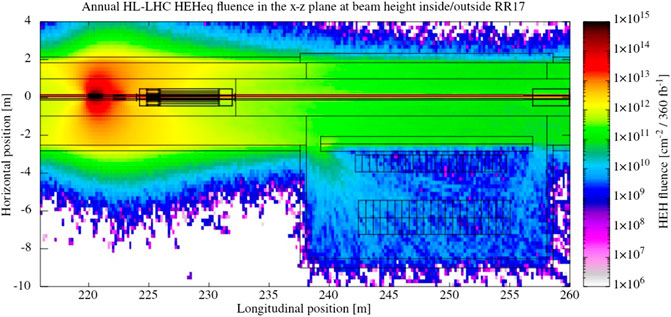}}
    \caption{(a) Comparison between FLUKA simulations and BLM TID measurements in 2018 in the LHC tunnel on the right side of the ATLAS experiment (from Ref.~\cite{bib:Na22LHC_TAN}) and (b) annual HL-LHC HEH-eq fluence simulated with FLUKA in the RR17 shielded alcove and the nearby LHC tunnel (from Ref.~\cite{bib:Ahdida2022}). }
    \label{fig:monitoring_calculation_radiation_levels}
\end{figure}

At the LHC, the key radiation quantities relevant for Radiation to Electronics (R2E) are monitored using a variety of complementary instrumentation. These include BLMs (already shown in Fig.~\ref{fig:LHC_BLMs}), RadMons (developed specifically for monitoring radiation effects on electronics, and shown in Fig.~\ref{fig:LHC_RadMons})~\cite{bib:FioreRadMon}, distributed optical-fibre dosimeters~\cite{bib:DiFrancescaFiberDosimetry}, radiophotoluminescence (RPL) passive dosimeters~\cite{bib:PrambergerRPL}, Timepix-based detectors~\cite{bib:PrelipceanTimepix3}, among others. Experimental measurements are systematically combined with Monte Carlo simulations, as illustrated in Fig.~\ref{fig:monitoring_calculation_radiation_levels}. This figure shows a benchmarking of FLUKA simulations against BLM measurements in the vicinity of the ATLAS experiment, as well as a two-dimensional map of the HEH-eq fluence simulated with FLUKA in a shielded alcove hosting electronics next to the LHC tunnel approximately 250~m from ATLAS.

In the mixed radiation fields characteristic of high-energy hadron accelerators, the total ionising dose (TID) and the HEH-eq fluence (together with other quantities used to characterise radiation levels) are generally well correlated. A commonly used rule of thumb is a conversion of approximately 1~Gy~$\approx 10^{9}$~HEH-eq/cm$^{2}$, although significant local variations can arise depending on the layout and shielding configuration. Quantitatively, radiation levels across different areas of the accelerator complex can span several orders of magnitude~\cite{bib:Bilko2023}, leading to markedly different risk profiles and corresponding requirements for electronic systems, as summarised in Table~\ref{tab:R2E_LHC_orders_of_magnitude}.

\begin{table}[h!]
\centering
\renewcommand{\arraystretch}{1.3}
\caption{Radiation levels at high-energy hadron accelerators and their effects on electronics.}
\begin{tabular}{ccp{8cm}}
\hline\hline
$\boldsymbol{\Phi_{\rm HEH\text{-}eq}}$ & \textbf{TID for 10 years } & \textbf{Impact on Electronics} \\
\textbf{(cm$^{-2}$ year$^{-1}$)} & \textbf{(Gy)} & \\ \hline
$10^{5}$  & $\ll 1$ &
Possible SEE impact for commercial systems with many units and very demanding availability and reliability requirements \\ \hline
$10^{7}$  & $< 1$ &
SEE impact for systems with multiple units and demanding availability and reliability requirements \\ \hline
$10^{9}$  & $10$ &
SEE mitigation (e.g.\ redundancy) at system level; cumulative effects can start to play a role \\ \hline
$10^{11}$ & $1~\text{kGy}$ &
SEE mitigation (e.g.\ redundancy) at system level, very challenging TID level for COTS \\ \hline
$10^{15}$ & $10~\text{MGy}$ &
Rad-hard by design ASICs \\ \hline\hline
\end{tabular}

\label{tab:R2E_LHC_orders_of_magnitude}
\end{table}

\section{Radiation Protection (RP) at hadron accelerators}
\label{sec:rp}

\begin{figure}[ht]
    \centering 
    \subfloat[\label{fig:RP_activation_FLUKA}]{
    \includegraphics[width=0.49\linewidth]{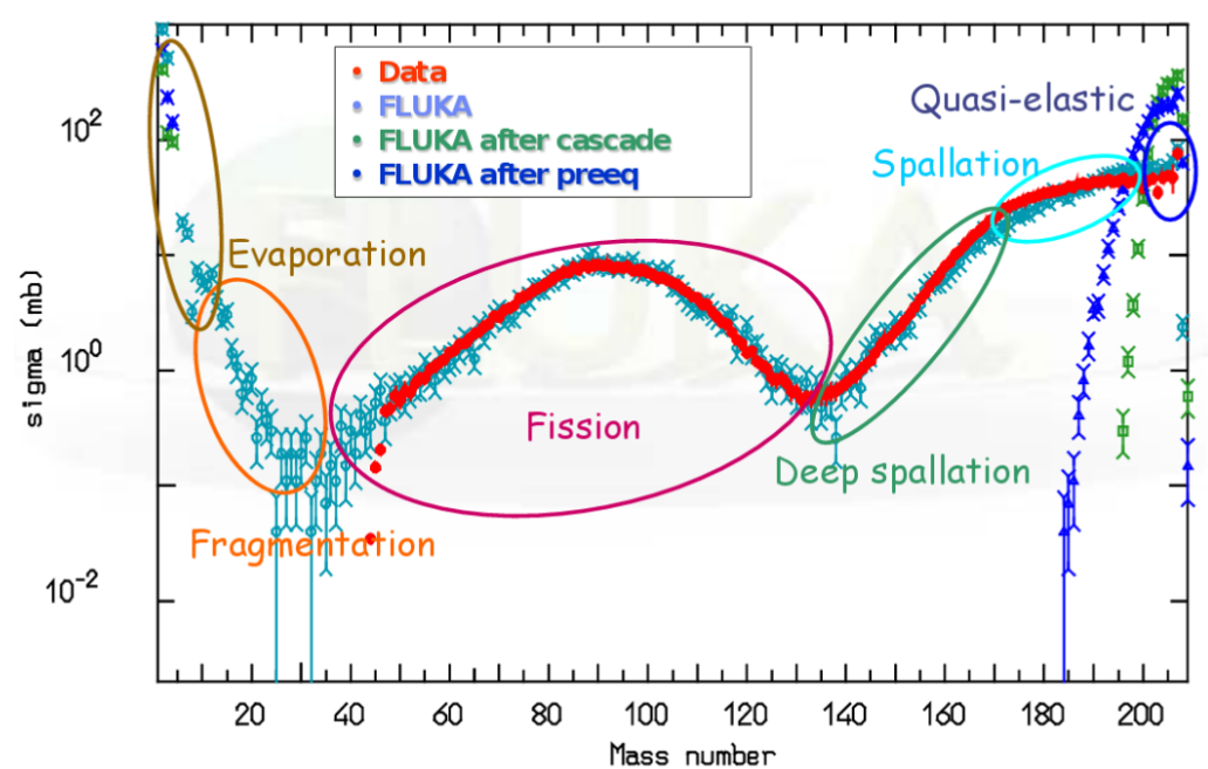}}
    \subfloat[\label{fig:RP_area_classification}]{
    \raisebox{0.15\height}{\includegraphics[width=0.495\linewidth]{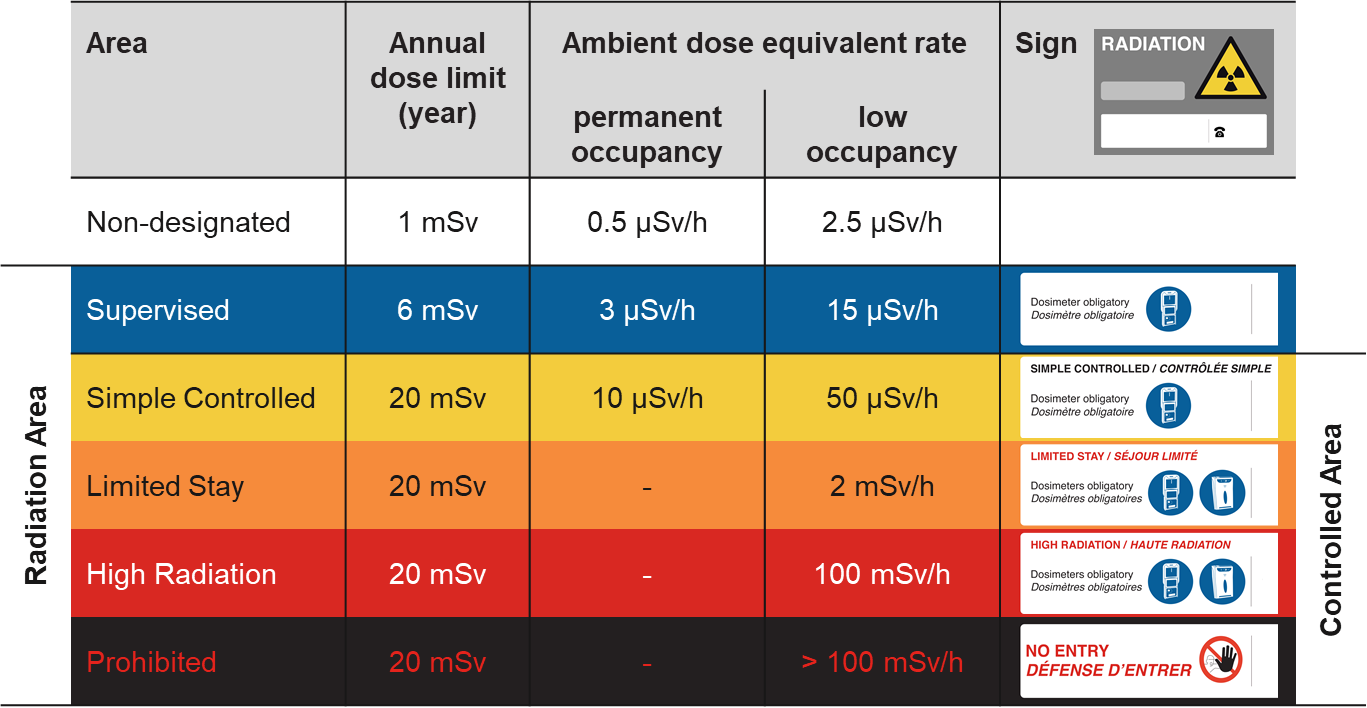}}} 
    \caption{(a) Residual nuclei production from 1~GeV protons on Lead simulated with FLUKA (from Ref.~\cite{bib:Battistoni2015}) and (b) classification of radiation areas at CERN based on the ambient dose equivalent. }
    \label{fig:RP_figures}
\end{figure}

As is well known, radiation is not only capable of causing damage to equipment and electronics, but it also poses a hazard to personnel and the public. For this reason, Radiation Protection (RP) must be ensured both against prompt radiation during accelerator operation and against residual radiation in experimental areas resulting from material activation. Nuclear reactions induced by beam losses at high-energy hadron accelerators produce a wide spectrum of residual nuclei, which may be either stable or radioactive and span a broad range of lifetimes. This is illustrated in Fig.~\ref{fig:RP_activation_FLUKA}, which shows the distribution of residual nuclei generated by 1~GeV protons impinging on a lead target, as simulated with FLUKA.

An important reference quantity for RP is the equivalent dose, expressed in sievert (Sv), which accounts for the different biological effects of various radiation types through particle-dependent weighting factors:
\begin{equation}
    H_T = \sum_R w_R \, D_{T,R},
\end{equation}
where $D_{T,R}$ is the absorbed dose in tissue $T$ from radiation type $R$, and $w_R$ is the corresponding radiation weighting factor. The absorbed dose $D_{T,R}$ is expressed in gray (Gy), as is the Total Ionising Dose (TID) defined in Eq.~(\ref{eqn:TID_def}). However, the two quantities are defined for different purposes: $D_{T,R}$ refers to energy deposition in biological tissue for radiation protection, while TID is defined for materials and electronic components. At CERN, the quantity used for operational purposes such as area monitoring and classification is the ambient dose equivalent, whose full definition is available in Ref.~\cite{bib:ForkelWirth2013} and which provides a conservative estimate of the equivalent dose suitable for routine measurements and simulations in accelerator environments.

%\begin{figure}[ht]
%    \centering 
%    \includegraphics[width=0.99\linewidth]{Figures/LHC_LSS1_Run3_ambient_dose_4months.png}
%    \caption{bla (a) (b) }
%    \label{fig:RP_LSS1_LS3_4months}
%\end{figure}

At the LHC, measurements and FLUKA simulations of the ambient dose equivalent (both prompt and residual after different cool-down times) are used to classify areas according to predefined limits, as illustrated in Fig.~\ref{fig:RP_area_classification}. Since activated materials can contain many different radioisotopes, each characterised by its own decay constant, the residual dose rate typically exhibits a superposition of exponential decays, with short-lived isotopes dominating shortly after beam stop and long-lived isotopes determining the dose at longer cool-down times. These classification criteria serve as the reference for defining access conditions and work regulations, which are crucial for planning interventions and maintenance activities.

\section{Beyond hadron accelerators: FCC-ee and muon colliders}
\label{sec:FCCee_mucol}

\subsection{FCC-ee}

When designing electron–positron accelerators such as the proposed FCC-ee at CERN~\cite{bib:FCCFeasibility}, the assessment of the consequences of beam losses represents a crucial task. Conventional loss mechanisms such as collision-driven losses, collimation losses, and beam–gas interactions still apply, with significant but not fundamental differences with respect to hadron accelerators. In addition, the light mass of electrons (0.511~MeV, nearly 2000 times smaller than the proton mass of 938~MeV) makes Synchrotron Radiation (SR)~\cite{bib:Hofmann2004_SR} a dominant design challenge. Indeed, the beam parameters at each of the FCC-ee operational modes (Z pole, WW threshold, ZH production, $t\bar{t}$) are constrained to limit the total SR power to 50~MW per beam. Although SR does not constitute a conventional beam loss (as the beam particles remain in circulation) it represents a source of radiation on the accelerator equipment, leading to analogous challenges and shielding requirements.

\begin{figure}[ht]
    \centering 
    \subfloat[\label{fig:FCCee_SR_spectra}]{
    \raisebox{0.05\height}{\includegraphics[width=0.38\linewidth]{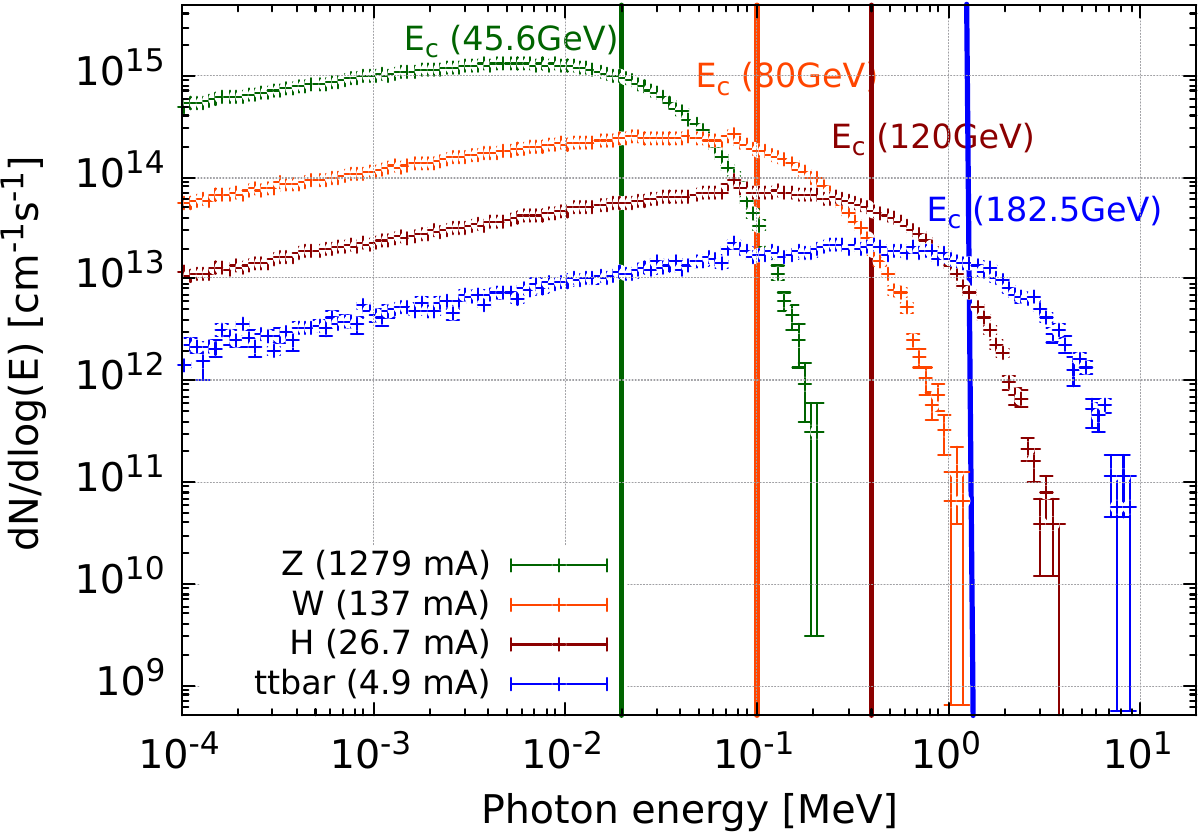}}}
    \subfloat[\label{fig:FCCee_arc_shielding}]{
    \includegraphics[width=0.6\linewidth]{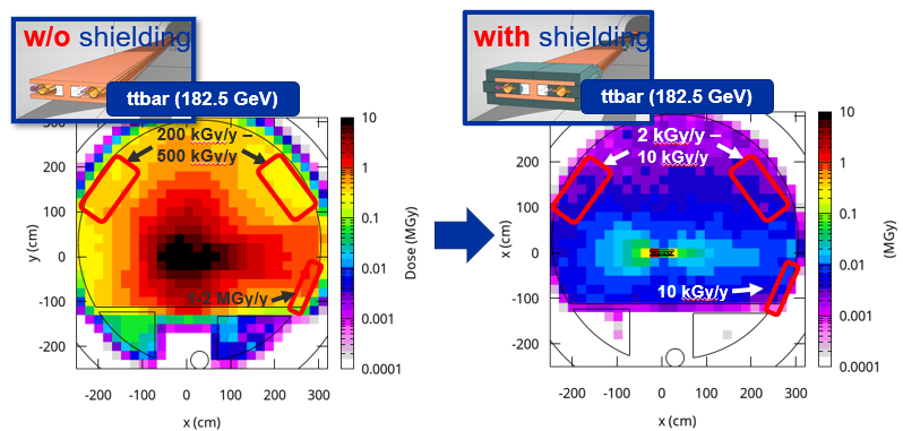}} 
    \caption{(a) Synchrotron Radiation spectra at the different operational energies of FCC-ee (from Ref.~\cite{bib:Humann2024}) and (b)~FLUKA simulation of the TID in a section of FCC-ee arc with and without dipole magnet shielding for $t\bar{t}$
 operation (from Ref.~\cite{bib:FCCFeasibility}). }
    \label{fig:FCCee}
\end{figure}

At the different beam energies, the arc dipoles emit significantly different SR spectra, as illustrated in Fig.~\ref{fig:FCCee_SR_spectra}. A key parameter characterising the spectrum is the critical energy $E_C$, defined as the photon energy that divides the total radiated power into two equal parts. The value of $E_C$ increases strongly with beam energy, producing harder photon spectra for the higher-energy operational modes.

Along the FCC-ee arcs, SR photons are intercepted by dedicated photon stoppers, concentrating the power absorption in discrete locations rather than along the magnets. For high-energy operation, where $E_C$ is of the order of the MeV, substantial shielding is required to maintain acceptable radiation levels in the tunnel, as shown by FLUKA maps of TID (defined in Eq.~(\ref{eqn:TID_def})) with and without shielding in Fig.~\ref{fig:FCCee_arc_shielding}. Current designs incorporate hundreds of kilograms of lead-based shielding per stopper, with multiple stoppers per dipole, making SR shielding a major technical and cost driver for the FCC-ee.

\subsection{Muon colliders}

Muon colliders offer clean multi-TeV lepton collisions without the synchrotron radiation limitations of electron–positron machines~\cite{bib:MuonCollider}. However, beam losses in muon colliders are intrinsically fast and unavoidable, being dominated by muon decay with a rest-frame lifetime of $\tau_{\rm rest} = 2.2\, \mu$s. Despite the~impact of time dilation, the lifetime in the laboratory frame for a 5-TeV muon beam remains limited to $\tau_{\rm lab} = \gamma_{\mu} \cdot \tau_{\rm rest} = (E_\mu/m_\mu)\cdot\tau_{\rm rest} \approx 0.1$~s, implying that essentially the entire beam is lost within a~fraction of a second through the decays $\mu^- \rightarrow e^- \bar{\nu_e} \nu_\mu$ and $\mu^+ \rightarrow e^+ \nu_e \bar{\nu_\mu}$.

\begin{figure}[ht]
    \centering 
    \subfloat[\label{fig:muon_collider_magnet_dose}]{
    \includegraphics[width=0.3\linewidth]{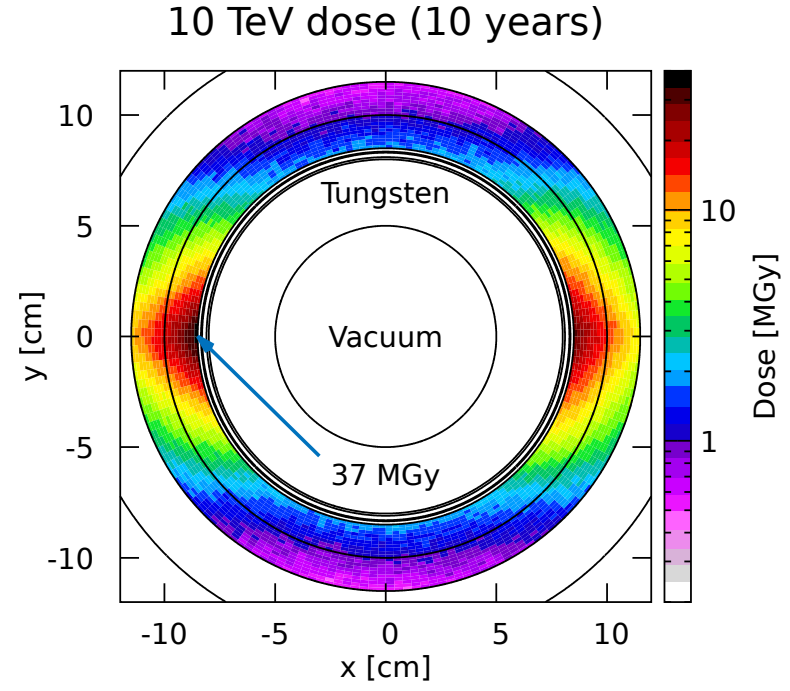}}
    \subfloat[\label{fig:muon_collider_neutrinos}]{
    \includegraphics[width=0.65\linewidth]{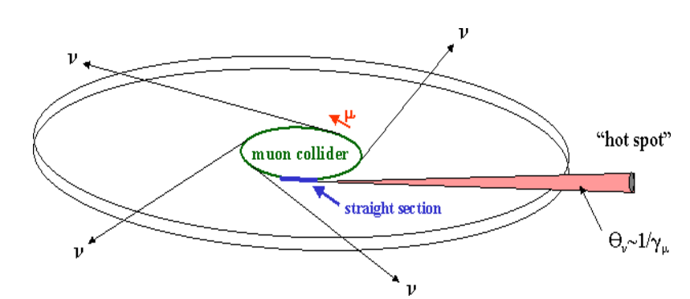}} 
    \caption{(a) FLUKA simulation of the dose deposited in the superconducting dipole magnets of a 10-TeV muon collider (from Ref.~\cite{bib:Calzolari2022Radiation}) and (b) illustration of the neutrino flux cone emitted by a straight section of a muon collider ring (from Ref.~\cite{bib:King2000MURINE}). }
    \label{fig:muoncollider}
\end{figure}

As a consequence, recent muon collider designs~\cite{bib:Accettura2024_IMCC} anticipate $\mathcal{O}(10^9)$ muon decays per meter per second during operation, distributed continuously along the collider ring. Each decay produces an~energetic electron or positron carrying, on average (but with a broad distribution), roughly one third of the~parent muon energy. These TeV-scale electrons initiate intense electromagnetic showers in the~surrounding accelerator components, leading to severe radiation and power deposition challenges. In particular, arc dipole magnets are exposed to very high radiation doses, reaching several tens of MGy in the superconducting coils over the collider lifetime, as illustrated by the FLUKA map in Fig.~\ref{fig:muon_collider_magnet_dose}, even with substantial tungsten shielding. In addition, decay products contribute significantly to experimental backgrounds at the interaction point. Each muon decay also produces two neutrinos, emitted in a narrow radial cone due to the large Lorentz boost, as shown in Fig.~\ref{fig:muon_collider_neutrinos}. Although neutrino interaction cross sections are small, they increase approximately linearly with energy, and the extremely high neutrino fluxes associated with multi-TeV muon beams may lead to Radiation Protection (RP) concerns at the Earth’s surface~\cite{bib:Manczak2025_NuDose}. This has motivated studies of neutrino flux mitigation strategies, such as deliberate orbit wobbling or site optimization, while simultaneously opening the possibility of exploiting muon colliders as intense sources for neutrino physics experiments.

\section{Conclusion}

Beam losses are an intrinsic aspect of the operation of high-energy and high-intensity accelerators and represent a central constraint for their safe and efficient performance. Building on a preceding lecture on particle–matter interaction principles~\cite{bib:LernerCAS2025interactions}, this article reviewed the main beam loss mechanisms in hadron accelerators, with particular emphasis on the LHC at CERN, and discussed their most critical consequences, including equipment damage, radiation effects on electronics, and radiation protection. The discussion was then extended to future facilities such as the FCC and muon colliders, where different types of processes become relevant. This non-exhaustive review highlights that beam losses and their consequences form a complex, multidisciplinary challenge, central to both the operation of current accelerators and the design of future ones.

\section*{Acknowledgements}
The material presented in this lecture, as well as in the previous one (Ref.~\cite{bib:LernerCAS2025interactions}), draws heavily on the expertise and resources available to the author as a member of the "Sources, Targets and Interactions" (STI) group at CERN. The author is particularly grateful to Anton Lechner, Francesco Cerutti, and Francesc Salvat Pujol, whose previous lectures provided both inspiration and valuable content for this work. He also extends special thanks to Francesco Cerutti for reviewing the proceedings of both lectures.


\begin{thebibliography}{99}

\bibitem{bib:LernerCAS2025interactions}
G.~Lerner,
``Particle--matter interactions,''
in \emph{CERN Accelerator School Proceedings: Intensity Limitations in Hadron Beams},
Borovets, Bulgaria, 2025.

\bibitem{bib:FCCFeasibility}
FCC Collaboration,
``Future Circular Collider Feasibility Study Report,''
\emph{Eur.\ Phys.\ J.\ C} \textbf{85} (2025),
Volumes~1--2,
CERN-FCC-2025-001,
arXiv:2505.00272.

\bibitem{bib:MuonCollider}
J.~P.~Delahaye \emph{et al.},
``Muon Colliders,''
arXiv:1901.06150 [physics.acc-ph].

\bibitem{bib:LHC}
L.~Evans and P.~Bryant (eds.),
``LHC Machine,''
\emph{JINST} \textbf{3} (2008) S08001.

\bibitem{bib:ATLAS}
ATLAS Collaboration,
``The ATLAS Experiment at the CERN Large Hadron Collider,''
\emph{JINST} \textbf{3} (2008) S08003.

\bibitem{bib:CMS}
CMS Collaboration,
``The CMS Experiment at the CERN LHC,''
\emph{JINST} \textbf{3} (2008) S08004.

\bibitem{bib:ALICE}
ALICE Collaboration,
``The ALICE Experiment at the CERN LHC,''
\emph{JINST} \textbf{3} (2008) S08002.

\bibitem{bib:LHCb}
LHCb Collaboration,
``The LHCb Detector at the LHC,''
\emph{JINST} \textbf{3} (2008) S08005.

\bibitem{bib:HLLHC-TDR}
O.~Aberle et al.,
``High‑Luminosity Large Hadron Collider (HL‑LHC): Technical Design Report,''
CERN Yellow Reports: Monographs, CERN‑2020‑010, CERN, Geneva, 2020,
doi:10.23731/CYRM‑2020‑0010,
\url{https://cds.cern.ch/record/2749422}. 

\bibitem{bib:FLUKA_website}
FLUKA Collaboration,
\emph{FLUKA: A Multi-Particle Transport Code},
\url{https://fluka.cern}.

\bibitem{bib:Ahdida2022}
C.~Ahdida \emph{et al.},
``New Capabilities of the FLUKA Multi-Purpose Code,''
\emph{Frontiers in Physics} \textbf{9}, 788253 (2022).

\bibitem{bib:Battistoni2015}
G.~Battistoni \emph{et al.},
``Overview of the FLUKA code,''
\emph{Annals of Nuclear Energy} \textbf{82}, 10--18 (2015).

\bibitem{bib:Redaelli2016}
S.~Redaelli,
``Beam Cleaning and Collimation Systems,''
arXiv:1608.03159 [physics.acc-ph],
CERN Yellow Report CERN‑2016‑002, pp.~403–437 (2016).

\bibitem{bib:BGC2024}
D.~Prelipcean et al.,
``Radiation levels from a beam gas curtain instrument at the LHC at CERN,''
in \emph{Proc.\ 15th International Particle Accelerator Conference (IPAC’24),} Nashville, TN, USA, 19–24 May 2024, paper THPG58,
JACoW/IPAC2024 (2024),
doi:10.18429/JACoW-IPAC2024-THPG58,
\url{https://cds.cern.ch/record/2912697}.

\bibitem{bib:Bilko2023}
K.~Biłko et al.,
``Radiation Environment in the Large Hadron Collider During the 2022 Restart and Related RHA Implications,''
\emph{IEEE Trans.\ Nucl.\ Sci.}, vol.~71, pp.~607--617, 2023,
doi:10.1109/TNS.2023.3328145. 

\bibitem{bib:Goddard825806}
B.~Goddard et al.,
\emph{TT40 Damage during 2004 High Intensity SPS Extraction},
CERN-AB-Note-2005-014, CERN, Geneva, Switzerland (2005),
\url{https://cds.cern.ch/record/825806}.

\bibitem{bib:Schoerling2023QuenchSlides}
D.~Schoerling, ``Heating -- quenches of superconducting magnets at the LHC,'' Training slides, CERN Accelerator School, 2023. Available: \url{https://indico.cern.ch/event/1254879/contributions/5331015/attachments/2687493/4662952/SS2023_1.pdf}.

\bibitem{bib:Holzer2005BLM}
E.~B.~Holzer \emph{et al.}, ``Beam loss monitoring system for the LHC,'' in 
\emph{IEEE Nuclear Science Symposium Conference Record}, Fajardo, PR, USA, 2005, pp.~1052--1056, doi:10.1109/NSSMIC.2005.1596433.


\bibitem{bib:RadiationToMaterials2023}
M.~Ferrari et al., ``“Radiation to Materials” at CERN,'' \emph{IEEE Trans.\ Nucl.\ Sci.}, vol.~70, no.~8, pp.~1580--1586, Aug.\ 2023, doi:10.1109/TNS.2023.3241785.

\bibitem{bib:LHCTripletTaskForce}
G.~Arduini et al., ``LHC Triplet Task Force Report,'' CERN Technical Report CERN-ACC-2023-0004, 2023. Available at \url{https://cds.cern.ch/record/2882512}.

\bibitem{bib:Uznanski2013REDW}
S.~Uznanski et al., 
``SEE and TID Test Results of Candidate Electronics for LHC Power Converter Control,''
in \emph{2013 IEEE Radiation Effects Data Workshop (REDW)}, San Francisco, CA, USA, 2013, pp. 1--5, doi: 10.1109/REDW.2013.6658206.

\bibitem{bib:Bitterling2016JINST}
O.~Bitterling et al., 
``Development of radiation tolerant components for the Quench Protection System at CERN,''
\emph{JINST}, vol.~11, C01032, 2016, doi: 10.1088/1748-0221/11/01/C01032.

\bibitem{bib:ASTM_E722}
ASTM International, 
\emph{ASTM E722-17: Standard Practice for Characterizing Neutron Fluence Spectra in Terms of an Equivalent Monoenergetic Neutron Fluence for Radiation Hardness Testing of Electronics}, West Conshohocken, PA, 2017, \url{https://www.astm.org/e0722-17.html}.

\bibitem{bib:Soderstrom2025TNS}
D.~S\"oderstr\"om et al.,
``Radiation Monitoring and Performance of Electronic Systems in High-Energy Accelerator Radiation Environments,''
\emph{IEEE Transactions on Nuclear Science}, vol.~72, no.~8, pp.~2490--2497, Aug.~2025, doi: 10.1109/TNS.2025.3534979.

\bibitem{bib:SpiezaRADMON}
G.~Spieza et al.,
``Operational Experience of the LHC Radiation Monitoring (RADMON) System,''
\emph{PoS} \textbf{RD11} (2011) 024,
doi:10.22323/1.143.0024.

\bibitem{bib:cecchettoNeutronSER}
M.~Cecchetto et al.,
``0.1--10 MeV Neutron Soft Error Rate in Accelerator and Atmospheric Environments,''
\emph{IEEE Trans.\ Nucl.\ Sci.} \textbf{68} (2021) no.~5, 873--883,
doi:10.1109/TNS.2021.3064666.

\bibitem{bib:Na22LHC_TAN}
S. Yang et al., ``$^{22}$Na activation level measurements of fused silica rods in the LHC target absorber for neutrals compared to simulations,''
\emph{Phys.\ Rev.\ Accel.\ Beams} \textbf{25} (2022) 091001,
doi:10.1103/PhysRevAccelBeams.25.091001.

\bibitem{bib:FioreRadMon}
S.~Fiore et al.,
``RadMon: a versatile, integrated radiation monitoring system for accelerators and experiments electronics at CERN,''
\emph{JINST} \textbf{20} (2025) C06015,
doi:10.1088/1748-0221/20/06/C06015.

\bibitem{bib:DiFrancescaFiberDosimetry}
D.~D.~Francesca et al.,
``Qualification and Calibration of Single-Mode Phosphosilicate Optical Fiber for Dosimetry at CERN,''
\emph{J.\ Lightwave Technol.} \textbf{37} (2019) no.~18, 4643--4649,
doi:10.1109/JLT.2019.2915510.

\bibitem{bib:PrambergerRPL}
D.~Pramberger et al.,
``Characterization of Radio-Photo-Luminescence (RPL) Dosimeters as Radiation Monitors in the CERN Accelerator Complex,''
\emph{IEEE Trans.\ Nucl.\ Sci.} \textbf{69} (2022) no.~7, 1618--1624,
doi:10.1109/TNS.2022.3174784.

\bibitem{bib:PrelipceanTimepix3}
D.~Prelipcean et al.,
``Towards a Timepix3 Radiation Monitor for the Accelerator Mixed Radiation Field: Characterisation with Protons and Alphas from 0.6~MeV to 5.6~MeV,''
\emph{Appl.\ Sci.} \textbf{14} (2024) no.~2, 624,
doi:10.3390/app14020624.

\bibitem{bib:ForkelWirth2013}
D.~Forkel-Wirth \textit{et al.}, ``Radiation protection at CERN,''
\textit{CAS -- CERN Accelerator School: High Power Hadron Machines},
2013, pp.~415--436, doi:~10.5170/CERN-2013-001.415.

\bibitem{bib:Hofmann2004_SR}
A.~Hofmann,
\textit{The Physics of Synchrotron Radiation},
Cambridge University Press, 2004.


\bibitem{bib:Humann2024}
B.~Humann et al., “Challenges and mitigation measures for synchrotron radiation on the FCC-ee arcs,” in Proc. 15th International Particle Accelerator Conference (IPAC’24), Nashville, TN, USA, May 2024, paper MOPG04, pp. 292–295, doi:10.18429/JACoW-IPAC2024-MOPG04.

\bibitem{bib:Accettura2024_IMCC}
C.~Accettura et al.,
``Interim report for the International Muon Collider Collaboration (IMCC),''
CERN Yellow Reports: Monographs, CERN-2024-002 (2024), 150 pp.,
doi:10.23731/CYRM-2024-002,
arXiv:2407.12450

\bibitem{bib:Calzolari2022Radiation}
D.~Calzolari et al.,
``Radiation load studies for superconducting dipole magnets in a 10 TeV muon collider,''
in \emph{Proc.\ of the 13th Int.\ Particle Accelerator Conf.\ (IPAC’22)}, Bangkok, Thailand, May 2022, pp.\ 1671--1674, 
doi:10.18429/JACoW-IPAC2022-WEPOST001. 

\bibitem{bib:King2000MURINE}
B.~J.~King,
``Mighty MURINEs: Neutrino physics at very high energy muon colliders,''
\emph{AIP Conf. Proc.} \textbf{530}, 142--164 (2000),
doi:10.1063/1.1361674.

\bibitem{bib:Manczak2025_NuDose}
J.~Manczak \textit{et al.},
``Refined FLUKA simulation model of neutrino-induced effective dose from a multi-TeV muon collider,''
in \emph{Proc.\ of the 16th Int.\ Particle Accelerator Conf.\ (IPAC’25)},
Taipei, Taiwan, June~2025, pp.\ TUPM076,
doi:10.18429/JACoW-IPAC2025-TUPM076.


\end{thebibliography}
\end{document}